\documentclass[letterpaper,11pt,onecolumn]{article}

\usepackage[]{epsf,epsfig, amsmath,amssymb,amsthm, latexsym, color,graphicx, xspace, url}

\usepackage{times,euscript, balance}

\usepackage{xspace}

\newtheorem{theorem}{Theorem}[section]

\newtheorem{observation}[theorem]{Observation}
\newtheorem{corollary}[theorem]{Corollary}
\newtheorem{definition}[theorem]{Definition}

\def\eps{{\varepsilon}}
\def\phi{{\varphi}}
\def\polylog{{\rm polylog}}

\def\A{\EuScript{A}}

\def\T{\EuScript{T}}

\def\V{\EuScript{V}}

\def\etal{\textit{et~al.}}

\def\bd{{\partial}}

\def\reals{{\mathbb R}}

\def\bd{{\partial}}

\def\xx{{\bf x}}

\newdimen\instindent
\def\institute#1{\gdef\@institute{#1}}

 \newfont{\affaddr}{phvr at 11pt}
 \newfont{\affaddrit}{phvro at 11pt} 

\usepackage{graphicx,floatflt,psfrag}

\begin{document}

\begin{titlepage}
\title{The Decision Tree Complexity for $k$-SUM is at most Nearly Quadratic\thanks{
    Work on this paper by Esther Ezra has been supported by NSF CAREER under grant CCF:AF 1553354.
    Work on this paper by Micha Sharir was supported by Grant 892/13 from the Israel Science Foundation,
    by Grant 2012/229 from the U.S.--Israel Binational Science Foundation, by the Israeli Centers of Research
    Excellence (I-CORE) program (Center No.~4/11), and by the Hermann
    Minkowski-MINERVA Center for Geometry at Tel Aviv University.}
}

\author{Esther Ezra
  \footnote{
    School of Mathematics,
    Georgia Institute of Technology, Atlanta, Georgia 30332, USA.
    \texttt{eezra3@math.gatech.edu}.
  }
  \and
  Micha Sharir
  \footnote{School of Computer Science, Tel Aviv University,
    Tel Aviv 69978 Israel and Courant Institute of Mathematical Sciences,
    New York University, New York, NY 10012, USA;
    \texttt{michas@post.tau.ac.il}.
  }
}

\maketitle

\begin{abstract}
  Following a recent improvement of Cardinal~\etal~\cite{CIO-16} on the complexity of a linear decision tree for $k$-SUM,
  resulting in $O(n^3 \log^3{n})$ linear queries, we present a further improvement to $O(n^2 \log^2{n})$ such queries.
\end{abstract}

\end{titlepage}


\section{Introduction}
\label{sec:intro}

\paragraph{Problem definition and the model.}


In this paper we study the \emph{$k$-SUM} problem, and the more general \emph{$k$-linear degeneracy testing ($k$-LDT)} problem.
We define them formally:

\begin{definition}[$k$-SUM]
  Given a point $\xx := (x_1, x_2, \ldots, x_n) \in {\reals}^n$ , decide whether there
  exist indices $i_1 < i_2 < \cdots < i_k$ such that $x_{i_{1}} + x_{i_{2}} + \ldots + x_{i_{k}} = 0$.
\end{definition}

\begin{definition}[$k$-LDT]
  Given a $k$-variate linear function $f(y_1, \ldots, y_k) = a_0 + \sum_{i=1}^k a_i y_i$, where $a_0, a_1, \ldots, a_k$
  are real coefficients, and a point $\xx := (x_1, x_2, \ldots, x_n) \in {\reals}^n$ , decide whether there
  exist indices $i_1 < i_2 < \cdots < i_k$ such that $f(x_{i_1}, x_{i_2}, \ldots, x_{i_k}) = 0$.
\end{definition}


By definition, $k$-SUM is a special case of $k$-LDT when we set $f(y_1, \ldots, y_k) = \sum_{i=1}^k y_i$.
Furthermore, the special case $k=3$ results in the so-called \emph{3-SUM} problem, which received considerable
attention in the past two decades, due to its implications to conditional lower bounds on the complexity of fundamental 
geometric problems; see~\cite{GO-95} and below for a list of such problems.
From now on we focus merely on the $k$-SUM problem, and show the (straightforward) extension to $k$-LDT only after we
present our algorithm and its analysis in Section~\ref{sec:algorithm}.
 
Following the approach in~\cite{CIO-16} for $k$-SUM (see also~\cite{AC-04, Erickson-99}), let $H$ be the collection 
of the ${n \choose k}$ hyperplanes $h$ of the form: $x_{i_1} + x_{i_2} + . . . + x_{i_k} = 0$. 
Then the $k$-SUM problem can be solved by locating the point $\xx$ in the \emph{arrangement} $\A(H)$ formed by those hyperplanes.
Specifically, this is done by a sequence of \emph{linear queries} of the form ``Does $\xx$ lie on, above, or below h?'', 
each of which is in fact a sign test, asking for the sign of $h(\xx)$, where $h(\cdot)$ is the linear expression defining $h$.

Our model is the \emph{$s$-linear decision tree}:
Solving an instance of the problem with input $\xx=(x_1,\ldots,x_n)$ is implemented as a search with $\xx$ in some tree $T$.
Each internal node $v$ of $T$ is assigned a linear function in the $n$ variables $x_1,\ldots,x_n$, with at most $s$ non-zero coefficients.
The outgoing edges from $v$ are labeled $<$, $>$, or $=$, indicating the branch to follow depending on the sign of the expression evaluated at $v$. 
Leaves are labeled ``YES'' or ``NO'', where ``YES'' means that we have managed to locate $\xx$ on a hyperplane of $H$,
and ``NO'' means that $\xx$ does not lie on any hyperplane.
To solve an instance of the problem, we begin at the root of $T$. At each node $v$ that we visit, we test the sign at $\xx$
of the linear function at $v$, and proceed along the outgoing edge labeled by the result of the test. We conduct this search until we reach a leaf,
and output its label ``YES'' or ``NO''.
At each internal node the test (which we also refer to as a \emph{linear query}) is assumed to cost one unit, 
and all other operations are assumed to incur no cost. 
Thus the length of the search path from the root to a leaf is the overall number of linear queries performed by the algorithm
on the given input, and is thus our measure for its cost. In other words, the worst-case complexity of the algorithm, 
in this model, is the maximum depth of its corresponding tree.
As in~\cite{CIO-16}, when $s=n$ (the maximum possible value for $s$), we refer to the model just as ``linear decision tree''.

\paragraph{Previous work.}

The $k$-SUM problem is a variant of the SUBSET-SUM problem, known to be NP-complete (for arbitrary $k$); however, its behavior as a 
function of $k$ has not yet been fully studied. 
Specifically, Erickson~\cite{Erickson-99} showed that the $k$-SUM problem can be solved in 
time $O((2n/k)^{\lceil{k/2}\rceil})$ for $k$ odd and $O((2n/k)^{k/2} \log{(n/k)})$ for $k$ even.
Moreover, he showed a nearly-tight lower bound of $\Omega((n/k^{k})^{\lceil{k/2}\rceil})$ in the $k$-linear decision tree model
(see~\cite{AC-04} for a more comprehensive overview of Erickson's result). 
Ailon and Chazelle~\cite{AC-04} slightly improved Erickson's lower bound,
and extended it to the $s$-linear decision tree model, where $s > k$, showing a lower bound of 
$\Omega\left((nk^{-3})^{\frac{2k - s}{2\lceil{(s-k+1)/2}\rceil} (1-\eps_k)}\right)$, where $\eps_k > 0$ tends to $0$ as $k$ goes to $\infty$. 
As stated in~\cite{AC-04}, in spite of the strength of this latter lower bound, it is not very informative if $s$ is not $O(k)$.
In particular, when $s$ is arbitrarily large (the case studied in this paper), one can no longer derive a lower bound of the form 
$n^{\Omega(k)}$. Indeed, Meiser's point-location mechanism~\cite{Meiser-93} implies that the depth of the (linear) decision tree 
in this case is only polynomial in $k$ and in $n$. 
In fact, Meyer auf der Heide~\cite{Meyer-84} showed an upper bound of $O(n^{4} \log{n})$ for the more general problem of $k$-LDT, and 
Cardinal~\etal~\cite{CIO-16} improved this bound\footnote{We note however that the bound in~\cite{CIO-16} applies to $k$-LDT with 
  only \emph{rational} coefficients.} to $O(n^{3} \polylog{n})$.
Concerning lower bounds in this model of computation, Dobkin and Lipton~\cite{DL-79} showed a lower bound of $\Omega(n\log{n})$
on the depth of the decision tree for $k$-LDT, see also~\cite{BenOr-83, SY-82} for more general non-linear models of computation. 

The case $k=3$ appears in various geometric problems, which are also known as \emph{$3SUM$-hard}. This includes problems as
testing whether there exist three collinear points in a given set of $n$ points, testing whether the union of $n$ given triangles 
covers the unit square, checking for polygon containment under translation, visibility among triangles in $3$-space, 
planar motion planning (under translations and rotations), translational motion planning in $3$-space, and more. 
This has been studied in the seminal work of Gajentaan and Overmars~\cite{GO-95}, who showed subquadratic reductions to $3$-SUM 
from many of these problems.
Over the last two decades the prevailing conjecture was thus that any algorithm for $3$-SUM requires $\Omega(n^2)$ time. 
Recently this has been refuted by Gr{\o}nlund and Pettie~\cite{GP-14}, who presented a subquadratic algorithm to solve $3$-SUM 
(see also the more recent work of Chan and Lewenstein~\cite{CL-15}, as well as those of Gold and Sharir~\cite{GS-15} and Freund~\cite{Freund-15}).
Furthermore, they showed that in the $(2k - 2)$-linear decision tree model, only $O(n^{k/2}\sqrt{\log {n}})$ queries 
are required for $k$ odd. In particular, when $k=3$ this bound is $O(n^{3/2} \sqrt{\log{n}})$. 
Very recently this bound has further been improved by Gold and Sharir~\cite{GS-15} to $O(n^{3/2})$, or, more generally,
to $O(n^{k/2})$ for arbitrary $k$, under the \emph{randomized} $(2k - 2)$-linear decision tree model. 
Note that in all these cases, the best known lower bound is just the standard $\Omega(n\log{n})$ bound, and closing the gap between 
this bound and the aforementioned upper bounds has still remained elusive.

\paragraph{Our result.}

Our main result is an improvement by an order of magnitude over the recent bound of Cardinal~\etal~\cite{CIO-16} 
on the complexity of a linear decision tree for $k$-SUM and $k$-LDT. Specifically, we show:
\begin{theorem}
  \label{thm:main}
  The complexity of $k$-SUM and $k$-LDT in the linear decision-tree model is $O(n^2\log^2n)$, where the constant of proportionality is linear in $k$.
\end{theorem}

Our analysis uses a variant of the approach in~\cite{CIO-16}, based on the point-location mechanism of Meiser~\cite{Meiser-93}, 
where we locate the input point $\xx$ in $\A(H)$ using a recursive algorithm that exploits and simulates locally
the construction of an \emph{$\eps$-cutting} of $\A(H)$. The main difference between 
the construction of \cite{CIO-16} and ours is that they use a bottom-vertex triangulation
on the cells in the arrangement of (a subset of) $H$, which partitions each cell into simplices. 
Since the ambient dimension is $n$, each simplex is defined (in general) by $\Theta(n^2)$ hyperplanes of $H$;
see, e.g.,~\cite{CS-89, SA-95} and below. 
On the other hand, in our construction we partition the cells of $\A(H)$ using the \emph{vertical decomposition}
technique~\cite{CEGS-91,SA-95}, where each cell that it produces is defined by only $O(n)$ hyperplanes. 
With a careful construction of the (prism-like) cell containing $\xx$,
this will eventually decrease the overall number of linear queries by an order of magnitude, with respect to the bound
obtained in~\cite{CIO-16}. 

\section{Preliminaries}
\label{sec:prelim}

\paragraph{Arrangements and vertical decomposition.}

Let $H$ be a collection of $n$ hyperplanes in ${\reals}^d$.
The \emph{vertical decomposition} $\V(H)$ of the arrangement $\A(H)$ 
is defined in the following recursive manner 
(see~\cite{CEGS-91,SA-95} for the general setup, and \cite{GHMS-95} for the case of hyperplanes in four dimensions).
Let the coordinate system be $x_1, x_2, \ldots, x_d$, and let $C$ be a cell in $\A(H)$.  
For each $(d-2)$-face $g$ on $\bd{C}$, we erect a $(d-1)$-dimensional \emph{vertical wall} passing through $g$ and confined
to $C$; that is, this is the union of all the maximal $x_d$-vertical line-segments that have one endpoint on $g$ and are contained in $C$.
The walls extend downwards (resp., upwards) from faces $g$ on the top side (resp., bottom side) of $C$.
This collection of walls subdivides $C$ into convex vertical prisms, each of which is bounded by (potentially many) 
vertical walls, and by two hyperplanes of $H$, one appearing on the bottom portion and one on the top portion of 
$\bd{C}$, referred to as the \emph{floor} and the \emph{ceiling} of the prism, respectively; in case $C$ is unbounded,
a prism may be bounded by just a single (floor or ceiling) hyperplane of $H$, and, in more extreme situations, can also be unbounded in both directions. 
More formally, this step is accomplished by projecting
the bottom and the top portions of $\bd{C}$ onto the hyperplane $x_d=0$, and by constructing the \emph{overlay} of these
two convex subdivisions. Each cell in the overlay, when lifted back to $\reals^d$ and intersected with $C$, becomes
one of the above prisms.

Note that after this step, the two bases (or the single base, in case the prism is unbounded) of a prism may have 
arbitrarily large complexity, or, more precisely, be bounded by arbitrarily many hyperplanes. 
(Each base, say the floor base, is a convex polyhedron in $\reals^{d-1}$, namely in the hyperplane $h$
containing it, bounded by at most $2n-2$ hyperplanes, where each such hyperplane is either an intersection of $h$ 
with another original hyperplane, or the vertical projection onto $h$ of an intersection of the corresponding ceiling hyperplane with some other hyperplane.) 
Our goal is to construct a decomposition of this kind so that each of its prisms is bounded by no more than $2d$ hyperplanes, independent of $n$.
To do so, we recurse with the construction at each base of each prism. Each recursive subproblem is now $(d-1)$-dimensional, 
within the hyperplane supporting the appropriate base.  

Specifically, after the first decomposition step described above, we project each prism just obtained onto the $x_1 x_2 \cdots x_{d-1}$-hyperplane,
obtaining a $(d-1)$-dimensional convex polyhedron $C'$, which we vertically decompose using a similar procedure.
That is, we now erect vertical walls within $C'$ from each $(d-3)$-face of $\bd{C'}$ in the $x_{d-1}$-direction.
These walls subdivide $C'$ into $x_{d-1}$-vertical prisms, each of which is bounded by (at most) two facets of $C'$ and some of the 
vertical walls. We keep projecting these prisms onto hyperplanes of lower dimensions, 
and produce the appropriate vertical walls. We stop the recursion as soon as we reach 
a one-dimensional instance, in which case all prisms projected from previous steps become line-segments, requiring no further decomposition.
We now backtrack, and lift the vertical walls (constructed in lower dimensions, over all iterations), one
dimension at a time,
ending up with $(d-1)$-dimensional walls in the original cell $C$; 
that is, a $(d-i)$-dimensional wall is ``stretched'' in directions $x_{d-i+2}, \ldots, x_{d}$ (applied in that order), 
for every $i=d, \ldots, 2$. 

Each of the final cells is a ``box-like'' prism, bounded by at most $2d$ hyperplanes. Of these, two are original hyperplanes,
two are hyperplanes supporting two $x_d$-vertical walls erected from some $(d-2)$-faces,
two are hyperplanes supporting two $x_{d-1}x_d$-vertical walls erected from some $(d-3)$-faces, and so on.

Note that each final prism is defined in terms of at most $2d$ original hyperplanes of $H$.
This follows by backwards induction on the dimension of the recursive instance. Initially, we have
two hyperplanes $h^-$, $h^+$, which contain the floor and ceiling of the prism, respectively. 
We intersect each of them with the remaining hyperplanes of $H$ (including the intersection $h^-\cap h^+$),
and project all these intersections onto the $(d-1)$-hyperplane $x_d=0$. Suppose inductively that, when we are at dimension $j$,
we already have a set $D_j$ of (at most) $2(d-j)$ original defining hyperplanes, and that each hyperplane in the 
current collection $H_j$ of $(j-1)$-hyperplanes is obtained by an interleaved sequence of intersections and projections, 
which involves some subset of defining hyperplanes and (at most) one additional original hyperplane.
We now choose a new floor and a new ceiling from among the hyperplanes in $H_j$, gaining two new defining hyperplanes 
(the unique ones that define the new floor and ceiling and are not in $D_j$). We add them to $D_j$ to form $D_{j-1}$,
intersect each of them with the other hyperplanes in $H_j$, and project all the resulting $(j-2)$-intersections onto
the $(j-1)$-hyperplane $x_j=0$, to obtain a new collection $H_{j-1}$ of $(j-2)$-hyperplanes. Clearly, the inductive 
properties that we assume carry over to the new sets $D_{j-1}$ and $H_{j-1}$.

We apply this recursive decomposition for each cell $C$ of $\A(H)$, and thereby obtain the entire vertical decomposition $\V(H)$.
We remark though that our algorithm does not explicitly construct the whole $\V(H)$. In fact, it will only construct the
prism containing the query point $\xx$, and even that will be done somewhat implicitly. The description given above,
while being constructive, is made only to define the relevant notions, and to set the infrastructure within which our algorithm will operate.

\paragraph{$\eps$-cuttings.}

Given a finite collection $H$ of hyperplanes in ${\reals}^d$, an \emph{$\eps$-cutting} for $H$
is a subdivision of space into prism-like cells, of the form just defined, 
that we simply refer to as prisms\footnote{Originally, these cells were taken to be simplices, 
  although both forms have been used in the literature by now.}, 
such that every cell is crossed by at most $\eps|H|$ elements of $H$, where $0 < \eps < 1$ is the parameter
of the cutting. This is a major tool for a variety of applications, including our own. 
It has been established and developed in a number of studies~\cite{CF-90, Clarkson-87, Mat-91}, 
and is based on the epsilon-net theorem of Haussler and Welzl~\cite{HW-87}. 

Specifically, we proceed as follows. We draw a random sample $R$ of 
$O\left(\frac{d_0}{\eps} \log{\frac{d_0}{\eps}}\right)$ hyperplanes (with an appropriate constant
of proportionality), where $d_0=2d$ is the (maximum) size of the \emph{defining set} of a prism, 
as discussed above\footnote{More generally, this is the \emph{primal shatter dimension} of the set system defined by $H$ and by the 
  subsets of $H$ meeting prisms. We do not formally define these notions here; see, e.g.,~\cite{Chaz-01, Har-Peled-11} for further details.}. 
Then we construct the arrangement $\A(R)$ of $R$ and its vertical decomposition $\V(R)$.
By the random sampling technique of Clarkson and Shor~\cite{Clarkson-87},
it is then guaranteed, with constant probability, that each prism meets at most $\eps|H|$ hyperplanes of $H$.

The preceding construction uses vertical decomposition. An alternative construction, for arrangements of
hyperplanes, is the \emph{bottom-vertex triangulation}, already mentioned above (see~\cite{SA-95}).
It has the advantage over vertical decomposition that the number of cells (simplices) that it produces is
known to be the best possible bound $\Theta(|H|^d)$ (where the constant of proportionality depends on $d$), 
but its major disadvantage for the analysis in this work is that the typical size of a defining set of a simplex is 
$d_0 = d(d+3)/2$, as opposed to the much smaller value $d_0=2d$ for vertical decomposition; see above, 
and also~\cite{AS_handbook-00} for more details. 
We thus obtain:
\begin{theorem}
  \label{thm:cutting}
  Given a finite set $H$ of hyperplanes in $d$-space, a random sample $R$ of 
  $O\left(\frac{d}{\eps} \log{\frac{d}{\eps}}\right)$ hyperplanes of $H$ (with an appropriate absolute constant
  of proportionality) guarantees, with constant probability, that each prism in the vertical 
  decomposition $\V(R)$ of $\A(R)$ meets at most $\eps|H|$ hyperplanes of $H$.
\end{theorem}

We remark that the method that we use here is not optimal, from a general perspective, in several aspects: 
First, it does not involve the refining second sampling stage of Chazelle and Friedman~\cite{CF-90} (and of others), 
which leads to a slight improvement in the number of cells. 
More significantly, in higher dimensions there are no sharp known bounds on the complexity of
vertical decomposition, even for arrangements of hyperplanes (see~\cite{CEGS-91,Koltun-04a} for the general case, and
\cite{GHMS-95,Koltun-04b} for the case of hyperplanes). Nevertheless, these issues are irrelevant to the technique employed
here (mainly because, as already mentioned, we are not going to construct the entire vertical decomposition),
and the coarser method just reviewed serves our purposes just right.

\section{The Algorithm}
\label{sec:algorithm}

\subsection{Algorithm outline}

The high-level approach of our algorithm can be regarded as an optimized variant of the algorithm 
of Cardinal~\etal~\cite{CIO-16}, which is based on the point-location mechanism of Meiser~\cite{Meiser-93}.
We choose $\eps > 0$ to be a constant, smaller than, say, $1/2$, and apply
the $\eps$-cutting machinery, as reviewed in Theorem~\ref{thm:cutting}.
For a given input point $\xx$, the algorithm proceeds as follows.
\begin{description}
\item{(i)} Construct a random sample $R$ of $r := O\left(\frac{n}{\eps} \log{\frac{n}{\eps}}\right)$ hyperplanes of $H$
(recall that in our application, the dimension of the underlying space is $n$).

\item{(ii)} Construct the prism $\tau=\tau_\xx$ of $\V(R)$ that contains the input point $\xx$.
(If at that step we detect a hyperplane containing $\xx$, we stop and return ``YES'').

\item{(iii)} Recurse on the subset of hyperplanes of $H$ that cross $\tau$;
as is common, we refer to this set as the \emph{conflict list} of $\tau$, and denote it by $CL(\tau)$.

\item{(iv)} Stop as soon as $\tau$ does not meet any other hyperplane in its interior.
Then return ``YES'' if $\xx$ lies on the floor or ceiling of $\tau$ (actually, this would have already been detected
in step (ii)), and ``NO'' otherwise.
\end{description}

\paragraph{Algorithm correctness.}
We emphasize that at each recursive step we construct the prism $\tau$ only with respect to the conflict list stored at its
parent cell $\tau_0$ (initially, $\tau_0 = {\reals}^d$ and $CL(\tau_0) = H$), implying that $\tau$ is not necessarily contained in $\tau_0$.
Still, this does not violate the search process, as the fact that $\xx$ lies in the interior of $\tau_0$ implies that (i) it can lie only on 
the hyperplanes in $CL(\tau_0)$ and on no other hyperplane, and, in particular, (ii) it does not lie on either the floor or ceiling of $\tau_0$,
as otherwise the search would have terminated already at $\tau_0$. 
Thus step (iv) of our algorithm produces the correct answer.
We also observe that the sequence of cells $\tau$ constructed in this manner are spatially in no particular relation to one another.
The invariant maintained by the construction is that the prism $\tau$ constructed at some step is a prism of the vertical decomposition of 
the conflict list of the prism $\tau_0$ constructed in the previous step.
As long as the sample $R$ is drawn from $CL(\tau_0)$, Theorem~\ref{thm:cutting} continues to hold, 
and the conflict list of $\tau$ contains only $\eps |CL(\tau_0)|$ hyperplanes of $CL(\tau_0)$, as required.

\paragraph{Constructing the prism of $\xx$.}

We next describe the details of implementing step (ii).
Since the overall complexity of a prism is exponential in the dimension $n$, we do not construct it explicitly.
Instead we only construct explicitly its at most $2n$ bounding hyperplanes, consisting of a floor and 
a ceiling (or only one of them in case the cell $C_{\xx}$ in $\A(R)$ containing $\xx$ is unbounded), 
and at most $2n-2$ vertical walls (we have strictly fewer than $2n-2$ walls in case the floor of $\tau$ intersects its ceiling,
or if this happens in any of the projections $\tau^{*}$ of $\tau$ in lower dimensions, or if the current subcell becomes unbounded at any of the
recursive steps---see below).
The cell $\tau$, as defined in step (ii), is then implicitly represented as the intersection of the halfspaces bounded
by these hyperplanes and containing $\xx$. Let $H_{\xx}$ denote this set of at most $2n$ hyperplanes.
From now on we assume, to simplify the presentation but without loss of generality, 
that $C_{\xx}$ is bounded, and that $\tau$ has exactly $2n-2$ vertical walls (and thus exactly $2n$ bounding hyperplanes).



We note that by construction all hyperplanes in $H$ are axis-parallel, meaning that they are not in general position,
and therefore their vertical decomposition is highly degenerate. In order to overcome this hurdle, we need to 
rotate the coordinate frame by, say, applying a sequence of $n-1$ infinitesimal rotations around the axes $x_2, x_3, \ldots, x_n$
(in order); see, e.g.,~\cite{VanArsdale-94} for details concerning representations of rotations in high dimensions and their associated 
transformation matrices.
We can thus assume from now on that no hyperplane in $H$ is axis-parallel.
Note that we also need to apply this rotation to the query point $\xx$. Observe that this does not violate our model of computation, in the
sense that the queries remain linear after the rotation. 
The only ``catch'' is that we can no longer control the number of nonzero coefficients in the linear tests; see a remark at the end of this section.

The following recursive algorithm constructs $H_{\xx}$.
Initially, we set $H_{\xx} := \emptyset$.
We first perform $r$ linear queries with $\xx$ and each of the hyperplanes of $R$, resulting in a sequence of $r$ 
output labels ``above''/``below''/``on'', where the latter 
implies that there is a positive solution to our instance of $k$-SUM, in which case we stop the entire procedure
and output ``YES''. We thus assume, without loss of generality, that all labels are ``above''/''below''. 
We next partition the set of the hyperplanes in $R$ according to their label, letting $R_{1}$ denote the set of
hyperplanes lying above $\xx$, and $R_{2}$ the set of hyperplanes below it.
We then identify the upper hyperplane $h_1 \in R_1$ and the lower hyperplane $h_2 \in R_2$ 
with shortest vertical distances from $\xx$. We do this by computing the minimum of these vertical distances, 
each of which is a linear expression in $\xx$, using $(|R_1|-1)+(|R_2|-1) < r$ additional comparisons.
The hyperplanes $h_1$ and $h_2$ contain the ceiling and the floor of $\tau$, respectively,
and we thus insert them into $H_{\xx}$.


In order to produce the hyperplanes containing the vertical walls of $\tau$, we recurse on the dimension $n$.
This process somewhat imitates the one producing the entire vertical decomposition of $C_{\xx}$ described above.
However, the challenges in the current construction are to build only the single prism containing $\xx$, 
to keep the representation implicit, and to do this efficiently.
 
We generate all pairwise intersections $h_1 \cap h$ and $h_2 \cap h$, for $h \in R$, $h \neq h_1$, $h \neq h_2$; 
note that due to our assumption that $\tau$ has exactly $2n$ bounding hyperplanes, we can ignore $h_1\cap h_2$,
and we do ignore it, merely to simplify the presentation.
We thus obtain two collections $G_1$, $G_2$ of $(n-2)$-dimensional flats, which we project onto
the $x_1\cdots x_{n-1}$-hyperplane. We next observe:

\begin{observation}
  \label{obs:unique_charge}
  For a fixed hyperplane $h \in R$, $h \neq h_1, h_2$,
  only one of $g_1 := h_1 \cap h$ or $g_2 := h_2 \cap h$ can appear on $\bd\tau$. 
\end{observation}

\noindent{\bf Proof.}
Consider $K_h := h\cap\tau$, which is a convex polytope of dimension at most $n-1$. 
It suffices to show that $K_h$ does not meet both $h_1$ and $h_2$.
Assume to the contrary that $K_h$ does intersect both $h_1$ and $h_2$. Take points $q_1\in K_h\cap h_1$ and $q_2\in K_h\cap h_2$,
and let $f$ be the smallest-dimensional face of $\tau$ that contains $q_1q_2$. Note that $f$ cannot be
the entire cell $\tau$, for then $q_1q_2$, and thus $h$, would cross the interior of $\tau$, contrary to its definition.
Moreover, because of the minimality of its dimension, $f$ must be contained in $h$. 
Indeed, if this were not the case, $h$ would have intersected $f$ at a lower-dimensional convex portion $f'$,
which necessarily contains $q_1q_2$. If $f'$ is a subface of $f$, we reach a contradiction to the minimality
of the dimension of $f$. Otherwise, $f$ meets both open halfspaces bounded by $h$. This however is impossible, 
because $h$ does not cross the interior of $\tau$, so one (open) side of $h$ lies fully outside $\tau$, and cannot meet $f$.
These contradictions establish the claim that $f$ must be contained in $h$.

Finally, since $q_1\in h_1$ and $q_2\in h_2$, $f$ must be a subface of some vertical wall(s). 
But then $h$ must be a vertical hyperplane (i.e., parallel to the $x_n$-direction), which we have ruled out
by the original rotation of the coordinate frame. This final contradiction completes the proof.
$\Box$

\noindent{\bf Remark:}
An important feature of the proof, which will be needed in the sequel, is that it also holds when $\tau$
is \emph{any} convex vertical prism, i.e., a prism with arbitrarily many vertical walls. The only crucial 
assumption is that no hyperplane in $R$ is vertical. We will rely on this property when we discuss the query
complexity---see below.


Observation~\ref{obs:unique_charge} implies that we can discard one of $g_1, g_2$ (we will later explain, in the analysis of the query complexity,
how we identify which of the two to discard), and thus the subset $G \subseteq G_1 \cup G_2$ of the surviving intersections consists of 
at most $|R|-2$ flats (of dimension $n-2$). 
We denote by $R^{(1)}$ the set of their projections onto the $x_1x_2\cdots x_{n-1}$-hyperplane.

We continue the construction recursively on $R^{(1)}$ in $n-1$ dimensions. 
That is, at the second iteration, we project $\xx$ onto $x_n=0$; let $\xx^{(1)}$ be the resulting point.
Then we locate the upper hyperplane $h_1^{(1)}$ and the lower hyperplane $h_2^{(1)}$ lying respectively 
above and below $\xx^{(1)}$ (now in the $x_{n-1}$-direction), with shortest vertical distances from $\xx^{(1)}$ 
(again, in the $x_{n-1}$-direction). 
We now erect vertical walls from $h_1^{(1)}$, $h_2^{(1)}$ in the $x_{n}$-direction,
obtaining a pair of $(n-1)$-dimensional walls (hyperplanes), which we insert into $H_{\xx}$.
We form the set $R^{(2)}$ of $(n-3)$-flats to be processed at the next iteration, by obtaining all pairwise 
intersections $h_1^{(1)} \cap h^{(1)}$ and $h_2^{(1)} \cap h^{(1)}$, for $h^{(1)} \in R^{(1)}$, by discarding one out of
each such pair (according to Observation~\ref{obs:unique_charge}, as will be explained later), and by projecting them onto the 
$x_1\cdots x_{n-2}$-hyperplane. Again we have $|R^{(2)}| < |R|$.

More generally, we proceed recursively in this manner, where at each step $i$, for $i=1,2,\ldots,n$,
we have a collection $R^{(i-1)}$ of fewer than $|R|$ $(n-i)$-hyperplanes, and a point $\xx^{(i-1)}$,
in the $x_1\cdots x_{n-i+1}$-hyperplane. We find the pair of hyperplanes that lie respectively
above and below $\xx^{(i-1)}$ in the $x_{n-i+1}$-direction, and are closest to $\xx^{(i-1)}$ in that direction, 
and then produce a set $R^{(i)}$ of fewer than $|R|$ $(n-i-1)$-hyperplanes in the $x_1\cdots x_{n-i}$-hyperplane.
We also project $\xx^{(i-1)}$ onto this hyperplane, thereby obtaining the next point $\xx^{(i)}$. 
The construction of $R^{(i)}$ is performed similarly to the way it is done in cases $i=1,2$, 
described above, and Observation~\ref{obs:unique_charge} continues to apply, so as to ensure 
that indeed $|R^{(i)}| < |R|$.

We stop when we reach $i = n$, in which case we are given a set of at most $|R|$ points on the real line, 
and we locate the two closest points to the final point $\xx^{(n)}$.

To complete the construction, we take each of the hyperplanes $h_1^{(i-1)}$, $h_2^{(i-1)}$, obtained at
each of the iterations $i = 2, \ldots, n$, and lift it ``vertically'' in all the remaining directions
$x_{n-i+2},\ldots,x_n$ (technically, we take the Cartesian product of these hyperplanes with $i-1$
suitable copies of $\reals$), and add the resulting $(n-1)$-hyperplanes in $\reals^n$ to $H_{\xx}$.


\paragraph{Constructing the conflict list of $\tau$.} 

In step (iii) of the algorithm we recurse on the conflict list $CL(\tau)$ of $\tau$, namely, the subset of
those hyperplanes of $H$ that cross $\tau$. In the ``full'', standard RAM model of computation, this is
a costly operation (recall that $H$ is a very large set). However, it costs nothing in our decision tree model. This follows by noting that 
the discrete representation of $\tau$ is independent of the actual coordinates of $\xx$. Specifically,
$\tau$ is the intersection of (at most) $2n$ halfspaces in $\reals^n$, namely, those containing $\xx$ and
bounded by the hyperplanes in $H_{\xx}$. Each of these hyperplanes is defined in terms of some subset
of the set $H_\tau$ of the (at most) $2n$ defining hyperplanes of $\tau$ in $H$ (see the definition of defining sets in
our discussion about $\eps$-cuttings in Section~\ref{sec:prelim}), via a sequence of operations, each of which is either 
(i) taking the intersection of some $h\in H_\tau$ with a previously constructed flat, or
(ii) projecting some flat one dimension down, or
(iii) lifting, in (one or more of) the remaining coordinate directions, a lower-dimensional flat to a hyperplane 
in $\reals^n$. (Note that an operation of type (iii) is vacuous algebraically---the equation of the flat within its ambient 
space is the equation of the lifted hyperplane in $\reals^n$; all the added coordinate variables have zero coefficients.)

Once we have constructed $\tau$, in this implicit manner, we take each hyperplane $h\in H$ and test whether it intersects $\tau$.
This is easy to do using linear programming (LP), by regarding $H_{\xx}$ as the set of constraints and the non-constant portion of $h$ 
as the objective function. We collect all the hyperplanes that cross $\tau$ into the desired conflict list $CL(\tau)$.  

Since none of these operations (obtaining the equations of the hyperplanes of $H_{\xx}$ and running the
LP-problems to construct $CL(\tau)$) depend on the concrete values of the coordinates of $\xx$, this
part of the algorithm incurs no cost in the decision tree model. 

\paragraph{The query complexity.}

Let $\bar{\tau}$ be the \emph{undecomposed} convex prism containing $\tau$. That is, this is the prism obtained in the first stage of the 
vertical decomposition process reviewed in Section~\ref{sec:prelim}, where are are given the cell $C_{\xx}$ containing $\xx$ in the arrangement 
$\A(R)$ and then produce the undecomposed prisms of $C$ by erecting vertical walls (confined to $C$) in the $x_n$-direction from each 
$(n-2)$-face on $\bd{C}$. As we recall, this decomposes $C_{\xx}$ into pairwise openly disjoint convex vertical prisms, each with a fixed 
(single-hyperplane) floor and a similar fixed ceiling, but with a potentially large number of vertical walls.
The prism $\bar{\tau}$ in that collection is the one that contains $\xx$. 
Using a similar notation as above, let $h_1, h_2 \in R$ be the hyperplanes containing the ceiling 
and the floor of $\bar{\tau}$, and assume, without loss of generality, that $h_1$ and $h_2$ do not intersect within the closure of $\bar{\tau}$. 
Let $G_1, G_2$ be the collections of $(n-2)$-dimensional flats obtained by all respective pairwise intersections $h_1 \cap h$, $h_2 \cap h$, 
for $h \in R \setminus \{h_1 \cup h_2\}$, and let $F_1, F_2$ be the collections of their corresponding liftings in the $x_n$-direction.
Note that each vertical wall of $\bar{\tau}$ is contained in one of the hyperplanes of $\{F_1 \cup F_2\}$. An easy but important property is that 
within a given cell $C_{\xx}$, the pair $(h_1, h_2)$ uniquely determines $\bar{\tau}$ (or, alternatively, any other prism determined by 
$(h_1, h_2)$ must lie outside $C_{\xx}$).

The remark following Observation~\ref{obs:unique_charge} implies the following corollary:
\begin{corollary}
  \label{cor:unique_charge2}
  For a fixed hyperplane $h \in R$, $h \neq h_1, h_2$,
  only one of $g_1 := h_1 \cap h$ or $g_2 := h_2 \cap h$ can appear on $\bd{\bar{\tau}}$. 
\end{corollary}
In other words, $\bar{\tau}$ contains at most $r-2$ vertical walls; let $F \subseteq \{F_1 \cup F_2\}$ be the set of hyperplanes containing
these walls. 
In order to retrieve $F$ during the query process, we prepare a data structure that stores, for each cell $C$ of $\A(R)$ and for each
pair $h_1, h_2$ of hyperplanes appearing on the top and bottom portions of $\bd{C}$, the list of the $r-2$ possible (hyperplanes containing the)
vertical walls associated with the corresponding undecomposed prisms $\bar{\tau}$.
%
The crucial observation about the construction of this data structure is that it does not depend on the query point $\xx$, and therefore 
incurs no cost in the decision tree model.

We now analyze the query complexity, beginning with the analysis of step (ii).

At each recursive step $i$ we are given a corresponding set $R^{(i-1)}$ of at most $r$ hyperplanes. 
We test each of them with $\xx^{(i-1)}$ in order 
to find the cell $C_{\xx^{(i-1)}}$ in $\A(R^{(i-1)})$ containing $\xx^{(i-1)}$, as well as to determine, for each hyperplane in $R^{(i-1)}$, whether it lies 
above or below $\xx^{(i-1)}$ in the $x_{n-i+1}$-direction. Overall, we perform at most $r$ linear queries at this step.

Finding the hyperplanes $h_1^{(i-1)}$ and $h_2^{(i-1)}$ that lie directly above and below $\xx^{(i-1)}$
takes at most $r$ additional linear queries (each of which compares the values of two linear expressions
in $\xx^{(i-1)}$, for two respective hyperplanes, as explained above). The generation of the pairwise intersections $g_1 = h_1^{(i-1)} \cap h$, 
(resp., $g_2 = h_2^{(i-1)} \cap h$) is independent of $\xx^{(i-1)}$, so it incurs no cost in our model.
This results in a collection of at most $2r$ hyperplanes, from which we form the sets $F_1$, $F_2$ of hyperplanes containing the 
corresponding vertical walls, and we retrieve the relevant subset $F = F^{(i-1)}$ of $r-2$ hyperplanes by querying the data structure described above, 
a step that costs nothing in our model.
We now project $F^{(i-1)}$ onto the $x_1\cdots x_{n-i}$-hyperplane, and thereby obtain the set $R^{(i)}$ with which we recruse 
in dimension $n-i$; we stop as soon as $i=n$.
The invariant maintained at each step of the recursion is that we move in the data structure from an undecomposed prism $\bar{\tau}$ 
in ${\reals}^{n-i+1}$ to a collection of at most $r$ hyperplanes in ${\reals}^{n-i}$. That is, initially we have a set of $r$ hyperplanes,
we then locate the prism $\bar{\tau}$ containing $\xx$ and retrieve the at most $r$ (or, more precisely, $r-2$) hyperplanes containing its 
vertical walls, which we project onto ${\reals}^{n-1}$ thereby obtaining a subset $R^{(1)}$ of at most $r$ hyperplanes in ${\reals}^{n-1}$. 
Since Corollary~\ref{cor:unique_charge2} continues to hold in any dimension $i \le n$,
it is easy to verify, using induction on the dimension, that the above property is maintained at each step of the recursion. 


Overall, this amounts to $O(r) = O\left(\frac{n}{\eps} \log{\frac{n}{\eps}}\right)$ linear queries at each recursive step, 
for a total of $O\left(\frac{n^2}{\eps} \log{\frac{n}{\eps}}\right)$ linear queries over all linearly many steps;
since $\eps$ is a constant fraction smaller than $1/2$, the actual bound is $O(n^2 \log{n})$.

The remaining steps of the algorithm do not depend on the query point $\xx$ (the tests in step (iv) are actually performed in step (ii)).
The recursion is powered by the $\eps$-cutting machinery, which guarantees that at each step we eliminate a constant 
fraction of the hyperplanes in $H$, so the algorithm terminates within $O(\log{|H|})$ steps. 
Therefore the overall number of linear queries is 
$$
O(n^2 \log{n}\log{|H|}) = O(kn^2\log^2n) .
$$


\paragraph{The $k$-LDT problem.}

We note that the only assumption that we need to make on the input hyperplanes $H$ is that none of them is axis-parallel, which
is accomplished by rotating the coordinate frame, as described above. Therefore our algorithm can be applied to collections of hyperplanes
with arbitrary real coefficients (as long as these hyperplanes are in general position). 
We thus conclude that the overall number of linear queries required to solve an instance of the $k$-LDT problem is $O(k n^2\log^2{n})$ as well.

This completes the proof of Theorem~\ref{thm:main}.
$\Box$

\noindent{\bf Remark:}
As already noted, in the $k$-SUM algorithm the number of non-zero coefficients in the resulting linear queries is arbitrarily large,
due to the rotation of the coordinate frame. 
However, even in an ideal scenario where rotations are not needed, and one can process the original hyperplanes (with only $k$ non-zero coefficients), 
this would still result in linear tests with arbitrarily many non-zero coefficients, as the number of non-zero coordinates is in general doubled in each of
the steps that recurse on the dimension.

\paragraph{Concluding remarks and open problems.}

In this paper we showed a nearly-quadratic upper bound on the number of queries in the linear decision tree model.
Being a significant improvement over previous works~\cite{CIO-16, Meyer-84}, this still leaves a gap of an order of
magnitude with respect to the lower bound $\Omega(n \log{n})$. Recall that for the special case of $k=3$ and $s=4$
(the number of allowed non-zero coefficients), the currently best known bound is $O(n^{3/2})$~\cite{GS-15, GP-14}, which strengthens the 
conjecture that for arbitrary large $s$ the actual bound should be close to linear. 

Our algorithm suggests an alternative machinery for point location in high dimensions.
Specifically, Meiser~\cite{Meiser-93} presented an algorithm for point location in an arrangement of $n$ hyperplanes in ${\reals}^d$,
with query time $O(d^{5} \log{n})$ and space $O(n^{d+\kappa})$, for $\kappa > 0$ arbitrary (see also the follow-up work of Liu~\cite{Liu-04}
for a slightly better space bound).\footnote{We note that a closer inspection of the analysis in~\cite{Meiser-93} 
  shows that the actual space complexity is $n^{O(d)}$, where the constant of proportionality in the exponent is greater than $2$.} 
Our mechanism applies vertical decomposition in such arrangements, resulting in a data structure whose overall space complexity of $n^{O(d)}$, 
where the constant of proportionality in the exponent is only slightly bigger than that of Meiser~\cite{Meiser-93}, and the query time is 
$O(d^{4} \log{n})$, improving Meiser's query time by an order of magnitude. 
Nevertheless, the major obstacle in our construction is that the constant of proportionality in the space complexity is double exponential.
We believe this is an artifact of the analysis and that the actual constant of proportionality should be only singly exponential, 
as in~\cite{Meiser-93}.
We discuss these details and present our considerations in Appendix~\ref{app:point_location}.

\paragraph{Acknowledgments.}
The authors would like to thank Shachar Lovett for many useful discussions.

\appendix

\section{A Point-Location Mechanism}
\label{app:point_location}

Formally, the point-location problem in arrangements of hyperplanes is defined as follows.
The input is an arrangement $\A(H)$ of a set $H$ of $n$ hyperplanes in ${\reals}^d$ (assumed without loss of generality 
to be in general position). 
Given a query point $q$ the goal is to find the cell in $\A(H)$ (or a face of lower dimension) that contains $q$. 

Our strategy in Section~\ref{sec:algorithm} leads to an alternative mechanism for point location in high dimensions.
Roughly speaking, we follow the approach of Meiser~\cite{Meiser-93}, but replace bottom-vertex triangulation with 
vertical decomposition.
In this section we present this data structure in detail and analyze its performance in the standard RAM model.

Following Meiser's approach, we uniquely represent each face $f$ of $\A(H)$ by its \emph{position vector} $pv(f)$,
where $pv(p) := (pv_1(p), pv_2(p), \ldots, pv_n(p))$ for a point $p$ in $d$-space is an $n$-dimensional vector indicating
for each coordinate $pv_i(p)$ the position of $p$ w.r.t. $h_i \in H$, that is, $pv_i(p) \in \{ '+', 0, '-' \}$ if $p$ lies either above, on, or
below $h_i$, respectively, $i=1, \ldots, n$. Note that this relation is well defined if none of the hyperplanes in $H$ is vertical, we also comment that
$pv(\cdot)$ is an equivalence relation whose equivalence classes are the faces $f$ of $\A(H)$ of all dimensions.
Thus in order to locate a query point $q$ in $\A(H)$ we in fact need to find its position vector in $\A(H)$.

\paragraph{Preprocessing.}

In order to reduce space complexity, we propose a slight modification to the approach in Section~\ref{sec:algorithm},  
which is inspired by the construction in~\cite{CEGS-91}. Preprocessing is performed with $\eps$-cuttings as follows.

Let $0 < \eps < 1$ be a parameter to be fixed shortly. We draw a random sample $R$ of 
$r := O\left(\frac{d}{\eps} \log{\frac{d}{\eps}}\right)$ hyperplanes of $H$.
Then we store all undecomposed prisms as follows. For each pair $h_1, h_2 \in R$ we store all prisms $\bar{\tau}$ whose ceiling and floor are 
$h_1$ and $h_2$, respectively, we assume without loss of generality that $h_1 \cap h_2$ does not meet within the closure of $\bar{\tau}$.
We note that the main modification of this structure w.r.t. the one in Section~\ref{sec:algorithm} is that in this case we do not build the
entire arrangement $\A(R)$, instead, we only store the pairs $(h_1, h_2)$ and their corresponding undecomposed prisms (each of which lies in 
a different cell of $\A(R)$). Therefore, in order to uniquely represent a prism $\bar{\tau}$, whose ceiling and floor are $h_1$ and $h_2$,  
we will store in it
(i) the position vector of $\bar{\tau}$ w.r.t. the $2r-2$ hyperplanes in $\{F_1 \cup F_2\}$ containing its vertical walls, 
where $F_1, F_2$ are defined as in Section~\ref{sec:algorithm}, 
and (ii) the subset $F \subseteq \{F_1 \cup F_2\}$ of the at most $r-2$ hyperplanes containing the actual vertical walls of $\bar{\tau}$
(refer once again to Section~\ref{sec:algorithm}).
We project the vertical hyperplanes in $F$ onto ${\reals}^{d-1}$, and 
build a similar structure in recursion; we stop as soon as $d=1$, in which case we obtain
a representation for all vertical prisms $\tau$ in the vertical decomposition $\V(R)$ of the arrangement $\A(R)$.
This construction yields an implicit representation for the prisms of $\V(R)$, where each cell $\tau$ is represented by the unique sequence of 
$d$ consecutive pairs, each of which consists of the two hyperplanes bounding $\tau$ in directions $x_d, x_{d-1}, \ldots, x_1$ (in order).

In order to make the point-location process efficient, we store the (representation of the) undecomposed prisms $\bar{\tau}$ in a tree-structure $\T$
consisting of $d$ levels, where at the bottom we store all prisms $\tau \in \V(R)$. Each non-leaf node of $\T$ has $O(r^2)$ children, one for each pair  
$h_1, h_2 \in R$, where the pairs are sorted lexicographically by the indices $(i,j)$ of the hyperplane pairs. Each internal node representing a pair 
$(h_1, h_2)$ stores the set $\{F_1 \cup F_2\}$ of the hyperplanes containing the $2r-2$ vertical walls, as defined above.
Then the resulting (undecomposed) prisms $\bar{\tau}$ are stored in a compressed TRIE of depth $2r-2$, ordered according to their position vector.  
Each prism stores the set $F$ as defined above. 
We now build the tree-structure $\T$ recursively in each undecomposed prism until we reach the bottom level.

For each such cell $\tau \in \V(R)$ we
(i) build its conflict list $CL(\tau)$ (we describe this step below), and note that, by Theorem~\ref{thm:cutting}, $CL(\tau) \le \eps|H|$,
and (ii) compute its position vector w.r.t. the hyperplanes in $H \setminus CL(\tau)$.
We apply this construction recursively with $CL(\tau)$ in each cell $\tau \in \A(R)$, until $|CL(\tau)|$ becomes smaller than $r$
(at an appropriate level), in which case we construct the vertical decomposition of the arrangement of the full set of leftover hyperplanes.
We thus obtain another tree structure (the ``$\eps$-cutting tree''), where at each bottom leaf we store prisms (of all dimensions $0, 1, \ldots, d$), 
each of which with its full position vector that uniquely refers to a face of $\A(H)$.

To recap we have two tree structures, primary and secondary, the first is the $\eps$-cutting tree and the latter is $\T$.

\paragraph{Point-location query.}

We describe how to obtain the prism $\tau \in \V(R)$ containing a query point $q$.
Once we locate $\tau$, we keep recursing in the primary tree structure with the hyperplanes in $CL(\tau)$ until we reach a leaf, 
in which case we obtain the full position vector of $q$ w.r.t. $\A(H)$.



We now describe the search in the secondary tree $\T$ (with input set $R$ of $r$ hyperplanes in ${\reals}^d$).
We note that at each level of the search, the construction of $\tau$ is implicit; 
and, without loss of generality, we assume that $\tau$ has exactly $2d$ bounding hyperplanes. 
We now proceed as follows. First, we identify the upper and lower hyperplanes $h_1, h_2$,  with shortest vertical distances from $q$. 
This involves $O(r)$ comparisons overall, where each comparison takes $O(d)$ time in the RAM model,
and thus $O(dr)$ time in total. Then we search the pair $(h_1, h_2)$ in $\T$, which costs $O(\log{r})$ time in the RAM model.
We then generate all pairwise intersections $h_1 \cap h$ and $h_2 \cap h$, for $h \in R$, $h \neq h_1, h_2$,
and obtain two collections $G_1$, $G_2$ of $(d-2)$-dimensional flats, which we stretch in the $x_d$-direction in order to obtain the
collections $F_1$, $F_2$ of hyperplanes containing the corresponding vertical walls. 
We then compute the position vector $pv(q)$ of $q$ w.r.t. $\{F_1 \cup F_2\}$, and then 
make a search with $pv(q)$ as a key in order to obtain the undecomposed prism $\bar{\tau}$ containing $q$ and the actual collection $F$, 
with which we continue (after projection onto ${\reals}^{d-1}$) at the next recursive step. 
This step too takes overall $O(dr)$ time in the RAM model, in particular, it includes the position vector search in the TRIE structure,
in which we walk along a path of length $2r-2$, until we reach at the prism $\bar{\tau}$ containing $q$. 
We then recurse in this manner on the dimension $d$, and stop as soon as $d=1$. 
Altogether, the running time is $O(dr)$ in each such step, for a total of $O(d^2r)$ time over the entire recursion.

\paragraph{The resulting recursion equations:}

We first bound the maximum storage requirement in $\T$.
Let $P_d(r)$ be the maximum number of prisms generated in a single secondary structure $\T$,
with an input set of $r$ hyperplanes in ${\reals}^d$,
observe that $P_1(r) = O(r)$ as at the bottom of the recurrence we just store the actual vertical decomposition of $\A(R)$,
and its complexity is linear when $d=1$. Then by the above discussion we have:

\begin{equation}
  \label{eq:P_d}
  P_d(r) \le  
  \begin{cases} 
    r^2 P_{d-1}(2r) & \text{if $ d > 1$,}
    \\
    O(r) &\text{otherwise.}
  \end{cases}
\end{equation}

Using induction on $d$ it is easy to verify that $P_d(r) = O(r^{2d-1})$, where the constant of proportionality is double exponential in $d$.
\footnote{We comment that we can stop the recursion already at $d=4$, and then apply the vertical decomposition in~\cite{Koltun-04a}, 
  in which case the bound on $P_d(r)$ is reduced to $O(r^{2d-4+\delta})$, for any $\delta > 0$, where the constant of proportionality 
  depends on $\delta$ and goes to $\infty$ as $\delta \rightarrow 0$.}
Since each prism $\bar{\tau}$ stores its position vector w.r.t. the $2r-2$ hyperplanes in $\{F_1 \cup F_2\}$, as well as the set $F$,
as defined above, this adds a factor of $O(r)$ to the total storage complexity, and thus, overall, the storage complexity is $O(r^{2d})$.

We next bound the space complexity of the primary tree structure.
Let $S(n)$ denote the maximum storage requirement for an input set of $n$ hyperplanes in ${\reals}^d$. Then by the above discussion we have:

\begin{equation}
  \label{eq:S_n}
  S(n) \le  
  \begin{cases} C_1 n r^{2d} + C_2 r^{2d-1} S(\eps n) & \text{if $ n > r$,}
    \\
    O(r^{2d}) &\text{otherwise,}
  \end{cases}
\end{equation}
where $C_1, C_2 > 0$ are two appropriate constants that depend double exponentially on $d$. 
In case $n > r$, the first summand in the bound of $S(n)$ is the overall complexity of the conflict lists (as well as the position vector)
over all prisms of $\V(R)$.

%

Next, let $Q(n)$ denote the maximum running time of a point-location query. We have:

\begin{equation}
  \label{eq:Q_n}
  Q(n) \le  
  \begin{cases} C_3 d^2 r + Q(\eps n) & \text{if $ n > r$,}
    \\
    O(dr) &\text{otherwise,}
  \end{cases}
\end{equation}
where $C_3 > 0$ is an appropriate constant.

We next need to make our choice for $\eps$. Let $\eps = 1/d$, in this case it is easy to verify that the solution
for Inequality~(\ref{eq:Q_n}) is $Q(n) = O(d^2 r \log_{d}{n}) = O(d^4 \log{n}/\log{d})$. Regarding the space complexity and the 
solution of Inequality~(\ref{eq:S_n}), one can show using induction on $n$ that $S(n) = O(n^{2d+\kappa})$, for any $\kappa > 2d$,
and thus $S(n) = n^{O(d)}$. We have thus shown:

\begin{theorem}
  Given a set $H$ of $n$ hyperplanes in ${\reals}^d$, there is a data structure to solve point-location queries in $O(d^4 \log{n}/\log{d})$ time, 
  with overall space complexity of $n^{O(d)}$.
\end{theorem}

\noindent{\bf Remark:}
We note that Meiser~\cite{Meiser-93} derived similar recursive inequalities, and a closer inspection of the analysis in~\cite{Meiser-93} shows that 
the bound $O(n^{d+\kappa})$ on the storage with $\kappa > 0$ arbitrarily small implies that the choice of the parameter $\eps$ should be
much smaller than $1/d$, in fact, $1/\eps$ needs to be super exponential in $d$ (we need to make a similar choice in our analysis 
in order to guarantee that the space complexity is only slightly larger than $O(n^{2d})$). Therefore, in order to keep the query time polynomial 
in $d$, $\kappa$ must be chosen sufficiently large. We also note that choosing $\eps$ to be a constant fraction improves Meiser's query time 
to $O(d^4 \polylog{d} \log{n})$, whereas our bound improves to $O(d^3\log{d} \log{n})$, this, however, considerably increases the space complexity.


\end{document}